# The accretion disk of SS 433?
## Remarks on recent observations of Brackett lines

M G Bowler

Department of physics, University of Oxford
Keble Road, Oxford OX1 3RH   (michael.bowler@physics.ox.ac.uk)

**Abstract**  Two sets of observations of the Galactic microquasar SS 433 in the Brackett series have recently been published. In this short note I point out two aspects that may have escaped the notice of the authors. The first is that the intermittent outbursts ~ 500 km $s^{-1}$ to red and blue in, for example, stationary $H_\alpha$ are now shown to be intermittent in the stationary Brackett series as well. These fast components have in the past been tentatively allocated to the accretion disk. The second is that in the absence of such outbursts the Brackett series [9] show the characteristics of a disk like structure rotating at ~ 200 km $s^{-1}$. In [9] this structure is claimed to be the accretion disk. I discuss how this not consistent with a large body of data – it matters because were it true the mass of the compact object would be $\lesssim$ 2 $M_\odot$.

**1. Introduction**
The microquasar SS 433 is unique in the Galaxy in that it emits jets of relatively cool matter fairly continuously at speeds approximately one quarter that of the speed of light, along an axis that precesses with a period of 162 days. The system is binary with a period of 13.08 days and the precession axis nods with a period of about 6 days [1]. The precession and nodding of the jet axis are also reflected in the line of sight speed of a wind very visible in $H_\alpha$, polar and blowing from the accretion disk [2]. The longest continuous sequence of stationary $H_\alpha$ spectra [2] shows not only this wind but also a pair of components characteristic of emission from a rotating disk like structure, the rotational speed being close to 200 km $s^{-1}$. This cannot be an accretion disk because this pair does not show orbital motion – yet the wind from the accretion disk does [2, 3].
 The first observations that suggest emission from the accretion disk itself are later in the same sequence as analysed in [2]. After over 30 days the stationary $H_\alpha$ line broadened rapidly, additional components appearing to both the red and the blue. If those components come from a rotating ring then the relevant rotational velocity is approximately 500 km $s^{-1}$. These fast components seem to speed up with time and as a pair show oscillations in redshift with period approximately 13 days; they could well be manifestations of the accretion disk [4, 5].
 Similar fast components have been detected in both $H_\alpha$ and $H_\beta$ in a series of opportunistic observations [6]. The observations are scattered and so is the separation of the red and blue components, very roughly 1200 km $s^{-1}$ but so scattered there is little hope of being able to detect orbital motion of the compact object and its disk, if that is indeed the source.

That brings us to Brackett series spectra. A sequence of observations in Br$\gamma$ covering a full orbital period was analysed in [7]. This line was broad and required in addition to a fast wind a pair of components corresponding to a disk rotation speed of 200 km $s^{-1}$ and another pair corresponding to a rotation speed of 500 km $s^{-1}$. The first pair was attributed to the circumbinary disk considered in [2] and the second pair to the accretion disk. It was suggested in [7] that the accretion disk is only visible intermittently at optical wavelengths, perhaps when clouds break, but that the clouds are transparent to the infrared Brackett series. It is puzzling that in [7] there is no detectable orbital motion of the 500 km $s^{-1}$ pair and for that matter no trace of orbital motion in the wind component. Thus any additional observations in the infrared are welcome.

The data in Fig. 2 of [8] show a single Br$\gamma$ spectrum. The profile is double peaked and broad; it might easily be a member of the family of spectra shown in [7]. The Brackett series spectra shown in [9] are very different. The 500 km $s^{-1}$ components are just not there at all.

## 2. The recent Brackett series data

The data of [8] were not taken with a view to examining Brackett spectra. As part of their VLTI/GRAVITY spectro-interferometry program the authors resolve the spatial structure of the SS 433 system on an interesting scale, examining the Br$\gamma$ lines from the two opposite jets and also the stationary Br$\gamma$ line. They infer that the emission regions in the jets decay with distance roughly exponentially on a scale of 1.7 $10^{14}$ cm and that the stationary Br$\gamma$ line is emitted from a region elongated along the direction of the jet axis and extending over more than $10^{14}$ cm. It is a reasonable speculation that the polar wind above the disk contributes to emission at these distances; at say 2000 km $s^{-1}$ a parcel of wind would have to remain active for perhaps 5 days. These results are both striking and tantalising, but of no direct concern here – the accretion disk must surely have an outer radius of the order of $10^{12}$ cm. The stationary Br$\gamma$ profile exhibited in Fig.2 of [8] shows that in 2016 July this line was broad, double peaked and similar to profiles in Fig.1 of [7], taken 10 years earlier.

The data of [9] comprise 5 sets of observations in the H and K bands. These were made at various times between 2014 May and 2015 July. In Fig.1 of [9] are illustrated spectra from 2015 27 May where the Brackett series lines march off into the distance from Br$\gamma$ to beyond Br20. Each line has the profile of a tower with two extreme horns, characteristic of emission from a rotating disk like region seen close to edge on. These profiles are very similar to those in Paschen series spectra, shown in Fig.2 of [10]. In neither [9] nor [10] are there any signs of the 500 km $s^{-1}$ red and blue components attributed to the accretion disk in [4,7]. The width of the Br$\gamma$ line in [9] is approximately 0.005 $\mu$m, to be contrasted with widths of ~0.01 $\mu$m in [7], some 9 years earlier and in [8], about a year later. (Detailed analysis of Brackett profiles in [9] was carried out only on Br12.)

## 3. Speculations on the accretion disk

It seems very likely that the ~500 km $s^{-1}$ components observed in Br$\gamma$ [7] and giving rise to the double peaked Br$\gamma$ profile in [8] are from essentially the same source as the optical outbursts in H$\alpha$ [4, 5], in He I [see 11] and in the H$\alpha$ and H$\beta$ spectra of [6]. The new Br$\gamma$ data show that even in the far infrared 500 km $s^{-1}$ components switch off for extended periods [9]. These fast components were OFF for the first 40 days of the 2004 data in [11] (see [2]) and then switched ON for over 30 days (to the end of that sequence of observations)[4, 5]. The Balmer series lines in [6] were acquired in 2006, two separate periods covering 65 days and (later) 40 days. The Br$\gamma$ spectra of [7], covering one orbital cycle, were taken during the period between the two sequences in [6]. In all these 2006 spectra, the 500 km $s^{-1}$ components were ON. Then in the data of [9] the 5 sets of observations between 2014 and 2015 show only narrow structures; the broad components were OFF, only to reappear a year later in the single spectrum of [8]. Thus the fast components exhibit extended periods in each of the ON/OFF states and [9] demonstrates that they disappear as readily in the far infrared as in the visible part of the spectrum. The speculation is that the source itself switches between ON and OFF states. During OFF periods the stationary lines are dominated by shapes characteristic of a circumbinary disk orbiting at ~ 200 km $s^{-1}$[2, 9].

It is of course tempting to speculate that the 500 km $s^{-1}$ red and blue components are radiated from an accretion disk rotating at roughly that speed, perhaps with some instability; the fast components have rather irregular speeds when ON (see [6]). The OFF periods might be terminated when some eruption of the donor star or instability in the outer regions of the disk occur. It should, however, be remembered that there is little direct evidence that these fast components originate in an accretion disk. Observationally the single orbital period of Br$\gamma$ in [7] seems to hold no memory of the orbital motion of the compact object and its disk, whereas the H$\alpha$ optical spectra certainly do [4, 3]. I know of no good reason why the red and blue components (and the broad component assigned to the wind) should so differ between Br$\gamma$ in [7] and the H$\alpha$ sequence of [3, 4, 5]. There are only 10 spectra, spread over one orbital period in [7]; [6] has demonstrated the degree of variation of these lines over a few days. A much longer sequence of observations might resolve the problem – unfortunately the new Br series data cannot.

If we suppose that two components, one to the red and the other to the blue and separated by ~ 1000 – 1200 km $s^{-1}$ are radiated episodically by material sharing the orbital motion of the compact object, can we suppose that the source is the outer regions of the accretion disk? This has been the supposition in the past [2, 4, 5, 7] and it could very well be sound. However, both a rotating ring and an expanding ring, radiating uniformly round the circumference, generate the same pattern of twin peaks to the red and blue. A dense radial wind has been detected in the equatorial plane through absorption troughs about 150 km $s^{-1}$ to the blue [1, 3] but it does not seem that these P Cygni profiles invariably appear with the fast red and blue peaks. I do not think there are any distinctive features that enable us to dismiss summarily the possibility of formation in a radially expanding region. This is in contrast to the data on the twin peaks separated by ~ 400 km $s^{-1}$; these exhibit some unmistakable signatures of a circumbinary rotating disk like region, reviewed in the following section. Assigning to the accretion disk the fast red and blue components that appear episodically may be attractive and plausible, but no more than that.

## 4. Circumbinary disk lines

The 10 Br$\gamma$ spectra reported in [7], taken over a single orbital period, have profiles showing the fast twin peaks separated by $\sim 1000$ km $s^{-1}$ and an inner pair separated by $\sim 400$ km $s^{-1}$. In the 5 scattered spectra of [9] the outer pair have disappeared and the inner pair remains. In the H$\alpha$ spectra displayed in Fig. 2 of [5] the sequence commences with only the inner pair present; that inner pair persists after the outer pair erupts over a few days, though less readily visible. Prior to the eruptions, twin peaked profiles very similar to the $\sim 400$ km $s^{-1}$ pair are found in other spectral lines in the data of [11], He I, O I and the Paschen series [12]. It is hardly plausible that the inner pair of Br$\gamma$ components of [7] and the (single) pair structure of the Brackett lines in [9] have a different origin.

   While [12] shows O I and a Paschen line on a single day and that a day when the profiles are highly symmetric, I have had no trouble in finding Paschen profiles that match very well the Br12 profiles in each spectrum of [9] at (approximately) the same orbital phases. The Paschen profiles evolve with orbital phase in very much the way that the He I profiles evolve; that sequence can be found in [11]. (The daily spectra for the O I and Paschen regions have not been published, but I have access to those data.) In short, I think that the pairs of Brackett lines separated by approximately 400 km $s^{-1}$ must have the same physical origin as the matching pairs in the emission lines H$\alpha$, the Paschen series, He I and O I. Collectively, these profiles do not admit an origin in an accretion disk and do display characteristics of a circumbinary disk. Below is a brief recapitulation of the evidence.

   First, over the thirty days of H$\alpha$ spectra to be found in [2], the red and blue components are separated by $\sim 400$ km $s^{-1}$ but show no trace of orbital motion of the compact object. At the same time, a broad H$\alpha$ component is identified as having a source in the wind above the disk, because the line of sight wind speed oscillates with the nodding period of the disk and also changes slowly with precession [2]. The centroid of this wind component shows the orbital motion with an amplitude of $\sim 130$ km $s^{-1}$ [3]. In H$\alpha$ the pair of components separated by $\sim 400$ km $s^{-1}$ is characteristic of a rotating ring or disk radiating with approximate azimuthal symmetry, but it cannot be from the accretion disk.

   Secondly, the profiles of He I emission lines at 6678, 7065 Å over the same period each show twin peaks separated by $\sim 400$ km $s^{-1}$; their amplitudes alternate with a period of 13 days. The pattern is most readily appreciated from Fig. 2 of [11]. In the convention where orbital phase is 0.5 when the compact object and its disk are eclipsing the companion, they are moving fastest towards us at orbital phase 0.25 and fastest to red at 0.75. The blue component is strong and the red weak at 0.25, the two components equally bright at 0.5 and the red is strong, blue weak at 0.75. This is exactly the pattern predicted for a hot spot caused by proximity of the compact object stimulating He I in a circumbinary disk or ring (compare Fig. 2 of [11] with the model calculation in Fig.4 of [13])). [A radially expanding region so excited would be bluest at orbital phase 0. An accretion disk would not be expected to shift the component with greatest intensity periodically from red to blue and back in the manner observed. The dominant aspect of the pattern would be two components of equal intensity together tracing the orbit of the compact object. (It is of course possible that this pattern would be modulated at certain orbital phases by obscuration.)]

The Paschen lines show the same He I pattern. The physics is explained in [15] and the same argument applies to the Brackett lines. The centroids of the Brackett lines in [9] would thus be expected to oscillate with a 13 day period – and apparently do, judging by Fig. 10 of [9]. That figure is not necessarily the signature of an accretion disk.

The first and second points above indicate that there exists a ring or disk rotating at $\sim 200$ km $s^{-1}$, that the data are collectively inconsistent with this being the accretion disk but carry signatures of a circumbinary disk. There is yet another indicator. The O I line at 8446 Å has a set of profiles very like the H$\alpha$ sequence through the first 30 days of the data of [11]. After the outbursts of an accelerated wind and the $\sim 500$ km $s^{-1}$ components, those O I profiles continue to follow the same pattern, that of a disk rotating at $\sim 200$ km $s^{-1}$. This line shows no sign of the outbursts in H$\alpha$ and He I. Its source is neither concealed nor disturbed by these outbursts linked to the compact object and the accretion disk; surely the disk radiating O I 8446 Å, unusual in that it is fluoresced by Ly$\beta$, must be circumbinary [14]?

## 5. Discussion and conclusions

Taken by themselves, the Brackett spectra in [9] could perfectly well be from an accretion disk, as claimed. They could equally well be from a circumbinary disk, if it be admitted that proximity of the compact object to the disk creates a hotspot [15]. There are simply no data in [9] adequate to distinguish between these possibilities. However, collectively the daily spectra in many different emission lines, taken over an extended period and comprising the data set of [11, 12], are not consistent with an accretion disk rotating at $\sim 200$ km $s^{-1}$. They are consistent with a circumbinary disk and appear to admit no other explanation.

In [9] it is claimed that the mass of the compact object must be less than 2.2 and probably less than 1.6 $M_\odot$. This claim assumes that the spectra in [9] are those of an accretion disk. Similarly the authors of [10] inferred a mass $\lesssim 1.4$ $M_\odot$ if their Paschen spectra were from the accretion disk but noted that if the source were rather an excretion disk then a system mass $\gtrsim 40$ $M_\odot$ is implied. This agrees with those studies that have produced the positive evidence for a circumbinary disk rotating at $\sim 200$ km $s^{-1}$ and against this motion being accretion. I do not see how the claims in [9] can be right.

If the spectral characteristics of a disk rotating at $\sim 200$ km $s^{-1}$ are not from the accretion disk, where then are spectra that are emitted from accretion disk material? The only candidate that I know of is the pair of broad red and blue components separated by $\approx 1000$ km $s^{-1}$, reported in the H$\alpha$ spectra of [4] and [5], H$\alpha$ and H$\beta$ of [6] and in the Br$\gamma$ line of [7]. There is evidence that this feature exhibits orbital motion [4, 5] but is it really the accretion disk as opposed to radial outbursts? Whatever the nature of the source, it is erratic and episodic even in the far infrared.